\documentclass[runningheads]{llncs}

\usepackage[utf8]{inputenc}

\usepackage[sort,nocompress]{cite}

\usepackage{booktabs} 

\usepackage{tikz}
\usetikzlibrary{positioning,shapes,calc,patterns}

\usepackage{pgfplots}
\pgfplotsset{compat=1.14}
\usepgfplotslibrary{statistics}

\usepackage{listings}
\usepackage{algorithm}
\usepackage[noend]{algpseudocode}

\usepackage{mathtools}
\usepackage{amsfonts}
\usepackage{amssymb}

\usepackage{glossaries}
\makeglossaries
\newacronym{smc}{SMC}{sequential Monte Carlo}
\newacronym{mcmc}{MCMC}{Markov chain Monte Carlo}
\newacronym{cps}{CPS}{continuation-passing style}
\newacronym{ppl}{PPL}{probabilistic programming language}

\usepackage{hyperref}

\newcommand{\ttt}{\texttt}

\newif\ifscratchpad
\scratchpadfalse

\newcounter{notecounter}

\newcommand{\term}{\mathbf{t}}
\newcommand{\val}{\mathbf{v}}
\newcommand{\expr}{\mathbf{e}}
\newcommand{\aval}{\mathbf{av}}
\newcommand{\cstr}{\mathbf{cstr}}
\newcommand{\set} {\mathbf{set}}

\usepackage{graphicx}
\begin{document}
\title{Automatic Alignment of Sequential Monte Carlo Inference in Higher-Order
\\ Probabilistic Programs}
\titlerunning{Alignment of Sequential Monte Carlo in Probabilistic Programming}
%
\author{Daniel Lundén\inst{1} \and
David Broman\inst{1} \and
Fredrik Ronquist\inst{2} \and \\
Lawrence M. Murray\inst{3}}
\institute{%
  KTH Royal Institute of Technology, Stockholm, Sweden \and
  Swedish Museum of Natural History, Stockholm, Sweden \and
  Uppsala University, Uppsala, Sweden
}
\maketitle              
\begin{abstract}
Probabilistic programming is a programming paradigm for expressing flexible
probabilistic models. Implementations of probabilistic programming languages
employ a variety of inference algorithms, where sequential Monte Carlo methods
are commonly used. A problem with current state-of-the-art
implementations using sequential Monte Carlo inference is the alignment of
program synchronization points. We propose a new static analysis approach based
on the 0-CFA algorithm for automatically aligning higher-order probabilistic
programs. We evaluate the automatic alignment on a phylogenetic model, showing
a significant decrease in runtime and increase in accuracy.

\end{abstract}
\section{Introduction}\label{sec:intro}

Probabilistic programming
\cite{gordon2014probabilistic,goodman2008church,wood2014a,dippl,murray2018delayed}
is a programming paradigm for expressing probabilistic models. A
\emph{\gls{ppl}} includes two constructs: one for \emph{sampling} from
probability distributions, and one for \emph{conditioning} on data. We use a
construct called \texttt{weight} for the latter, which simply adds its argument
to a \emph{logarithmic\footnote{This is commonly done for numerical
stability.} weight} attached to the current execution. One motivation for using
probabilistic programming is greater \emph{expressive power} compared to
classical approaches to probabilistic modeling, such as Bayesian networks. This
increase in expressive power comes from two properties: \emph{stochastic
branching}, i.e. that control flow can depend on randomness, and
\emph{recursion}. A \gls{ppl} with these two properties is called a
\emph{universal} \gls{ppl}~\cite{goodman2008church}.

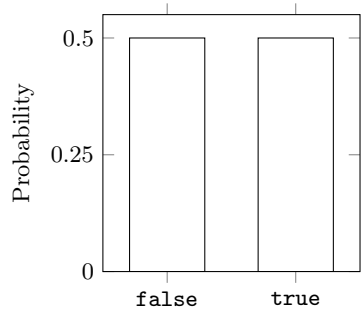
\begin{figure}[tb]
  \centering
  \begin{tabular}{c}
    \begin{lstlisting}
weight(5)
if flip() then {
  weight(10)
  weight(85)
  false
} else {
  weight(95)
  true
}
    \end{lstlisting}
  \end{tabular} \qquad
  \begin{tikzpicture}[baseline=(current bounding box.center)]
    \begin{axis}[
      ybar, ymin=0,
      bar width=10mm,
      width=5cm, height=5cm, enlarge x limits=0.5,
      ylabel={Probability},
      ytick={0,0.25,0.5},
      xtick=data,
      xticklabels={\texttt{false},\texttt{true}}
      ]
      \addplot [black] coordinates {(0,0.5) (1,0.5)};
    \end{axis}
  \end{tikzpicture}
  \caption{%
    A toy example illustrating when resampling can be problematic.  The example
    is written in our own functional, higher-order, \gls{ppl} (under
    development). The function \texttt{flip} represents a coin flip.  The bar
    plot shows the true distribution encoded by the example---that is, on
    average, there should be equally many executions resulting in \texttt{true}
    as executions resulting in \texttt{false} (with the constant weight of 100).
  }
  \label{fig:introex}
\end{figure}

The most important component of a \gls{ppl} is its \emph{inference algorithm},
which is loosely analogous to the execution semantics of ordinary programming
languages. \emph{\Gls{smc}} methods \cite{liu1998sequential} are commonly used
as such inference algorithms \cite{dippl,wood2014a,murray2018delayed}. They
perform inference by executing a number of instances of a probabilistic program
in parallel, pausing the executions when they encounter a conditioning on data.
When all executions have been paused, the algorithm looks at the weights of the
different executions given the data, and \emph{resamples} the set of executions
proportional to these weights. That is, more probable executions are
replicated, and less probable executions are discarded. This process repeats
until the program has reached its end. There are, however, problems with this
approach. The toy program in Fig.~\ref{fig:introex} encodes a probability
distribution over booleans using a stochastich branch. The different executions
in \gls{smc} inference for the program will encounter a different number of
calls to \texttt{weight}, either three or two.  Furthermore, they will not
always \emph{align} at the same \texttt{weight} statements simultaneously---it
is possible that one execution can pause at line 3, while another pauses at
line 7. In Fig.~\ref{fig:introex}, if we are only running a moderate number of
executions in total (say 10\,000), with overwhelmingly high probability, all
executions ending up at line 3 will be discarded; this is because of their low
weight ($e^{5+10}$) relative to the other weight at line 7 ($e^{5+95}$). We can
clearly see, however, that in the end both branches should be equally weighted.

The problem illustrated in Fig.~\ref{fig:introex} is not handled optimally by a
direct implementation of \gls{smc}. Such implementations are, for instance,
available in WebPPL \cite{dippl} and Anglican \cite{wood2014a}. When performing
\gls{smc} inference on an equivalent program in WebPPL, the algorithm performs
inference without any visible errors, but only returns \texttt{true}. In
Anglican, an error is given at runtime, stating that some
\emph{observes}\footnote{Anglican uses a different construct for conditioning
on data called \texttt{observe}.} are not global. This error is given for all
programs where different executions do not have the same number of calls to
\texttt{weight}.

It is possible for users to \emph{manually} align unaligned programs, taking
care to only place calls to \texttt{weight} where they are aligned. However,
for larger programs, manual alignment can become an error-prone process, and a
nuisance for the programmer. In this paper, we propose an \emph{automatic}
solution for aligning higher-order probabilistic programs using static
analysis. The static analysis is used to find all \emph{dynamic} terms in a
program---that is, terms that may be reached from within a stochastic branch.
In Fig.~\ref{fig:introex}, all terms within both branches of the \texttt{if}
expression are dynamic, since the condition is random. In particular, the calls
to \texttt{weight} on lines 3, 4, and 7 are dynamic, and hence unaligned. The
call to \texttt{weight} on line 1 is not dynamic, however, and is therefore
aligned. By identifying all unaligned \texttt{weight} calls, we can handle
these specially when running \gls{smc}, making the \gls{smc} inference
\emph{aligned}. The contributions are:
\begin{itemize}
  \item
    A static analysis algorithm, based on 0-CFA \cite{shivers1988control,shivers1991control},
    for discovering dynamic terms in higher-order probabilistic programs
    (Section~\ref{sec:disc}).
  \item
    An application of the above analysis, where the resulting dynamic
    terms are used to automatically align \gls{smc}
    inference for higher-order probabilistic programs (Section~\ref{sec:utilize}).
  \item
    An evaluation of our automatic alignment approach for \gls{smc} inference,
    compared to the unaligned \gls{smc} implementation, as found in
    WebPPL\footnote{We compare to the \gls{smc} algorihm found in WebPPL, since
    the Anglican \gls{smc} algorithm does not handle unaligned programs.}.
    This evaluation is performed through a case study on a model from
    phylogenetics (Section~\ref{sec:case}).
\end{itemize}
Before describing our contributions in detail, Section~\ref{sec:prelim} will
provide some necessary background.

\section{Preliminaries}\label{sec:prelim}
In this section, we give a brief introduction to a classical \gls{smc} method
for Bayesian networks. This background is needed to understand the inference
semantics of the \gls{ppl} presented in the later sections.


\paragraph{Bayesian networks.}
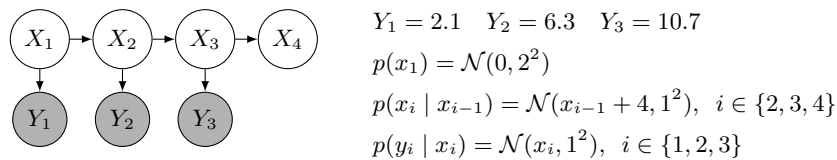
\begin{figure}[b]
  \centering
    \begin{tikzpicture}[node distance = 3mm,
      baseline=(current bounding box.center)]
      \node[draw,circle] (X1) {$X_1$};
      \node[draw,circle,right=of X1] (X2) {$X_2$};
      \node[draw,circle,right=of X2] (X3) {$X_3$};
      \node[draw,circle,right=of X3] (X4) {$X_4$};
      \node[fill=black!30,draw,circle,below=of X1] (Y1) {$Y_1$};
      \node[fill=black!30,draw,circle,below=of X2] (Y2) {$Y_2$};
      \node[fill=black!30,draw,circle,below=of X3] (Y3) {$Y_3$};

      \draw[-latex] (X1) -- (X2);
      \draw[-latex] (X2) -- (X3);
      \draw[-latex] (X3) -- (X4);
      \draw[-latex] (X1) -- (Y1);
      \draw[-latex] (X2) -- (Y2);
      \draw[-latex] (X3) -- (Y3);
    \end{tikzpicture}
    \quad
    \begin{minipage}{0.55\textwidth}
      \[
        \begin{aligned}
          & Y_1 = 2.1 \quad Y_2 = 6.3 \quad Y_3 = 10.7 \\
          & p(x_1) = \mathcal{N}(0,2^2) \\
          & p(x_i \mid x_{i-1}) = \mathcal{N}(x_{i-1} + 4,1^2),
          \enspace i \in \{ 2, 3, 4 \} \\
          & p(y_i \mid x_{i}) = \mathcal{N}(x_{i}, 1^2),
          \enspace i \in \{ 1, 2, 3 \}
        \end{aligned}
      \]
    \end{minipage}
    \caption{%
      A Bayesian network representation of a simple linear Gaussian state space
      model. The symbol $\mathcal{N}$ is a notation for the ubiquitous normal
      distribution.
    }
  \label{fig:baynet}
\end{figure}
A Bayesian network~\cite{pearl1985bayesian} is a \emph{directed acyclic graph}
where the vertices are \emph{random variables} and the edges direct
dependencies between them. An example of a Bayesian network is
given in Fig.~\ref{fig:baynet}. The random variables $X_i$ are the exact
positions of some moving object at time $i$.  The random variables $Y_1$,
$Y_2$, and $Y_3$ are noisy observations of the positions with values given in
the figure (shaded in the graph).
For more details on probability theory and Bayesian networks, see e.g., Bishop
\cite{bishop2006pattern}.

\begin{figure}[tb]
  \centering
    \begin{tikzpicture}[node distance = 1mm,minimum width = 2.4cm]
      \node[draw,rounded rectangle,                fill=black!0
        ] (S11) {$X_1 \approx -2.5$};
      \node[draw,rounded rectangle,below = of S11, fill=black!46
        ] (S12) {$X_1 \approx 4.7$};
      \node[draw,rounded rectangle,below = of S12, fill=black!54
        ] (S13) {$X_1 \approx 4.6$};
      \node[draw,rounded rectangle,below = of S13, fill=black!0
        ] (S14) {$X_1 \approx -3.2$};
      \node[draw,rounded rectangle,below = of S14, fill=black!0
        ] (S15) {$X_1 \approx -2.8$};
      \node[draw,rounded rectangle,right = 1cm of S11, fill=black!10
        ] (S21) {$X_2 \approx 8.9$};
      \node[draw,rounded rectangle,below = of S21, fill=black!46
        ] (S22) {$X_2 \approx 8.0$};
      \node[draw,rounded rectangle,below = of S22, fill=black!4
        ] (S23) {$X_2 \approx 9.2$};
      \node[draw,rounded rectangle,below = of S23, fill=black!44
        ] (S24) {$X_2 \approx 8.1$};
      \node[draw,rounded rectangle,below = of S24, fill=black!0
        ] (S25) {$X_2 \approx 9.9$};
      \node[draw,rounded rectangle,right = 1cm of S21, fill=black!30
        ] (S31) {$X_3 \approx 11.5$};
      \node[draw,rounded rectangle,below = of S31, fill=black!11
        ] (S32) {$X_3 \approx 12.3$};
      \node[draw,rounded rectangle,below = of S32, fill=black!21
        ] (S33) {$X_3 \approx 11.9$};
      \node[draw,rounded rectangle,below = of S33, fill=black!15
        ] (S34) {$X_3 \approx 12.1$};
      \node[draw,rounded rectangle,below = of S34, fill=black!23
        ] (S35) {$X_3 \approx 11.8$};
      \node[draw,rounded rectangle,right = 1cm of S31, fill=black!20
        ] (S41) {$X_4 \approx 15.9$};
      \node[draw,rounded rectangle,below = of S41, fill=black!20
        ] (S42) {$X_4 \approx 16.1$};
      \node[draw,rounded rectangle,below = of S42, fill=black!20
        ] (S43) {$X_4 \approx 15.4$};
      \node[draw,rounded rectangle,below = of S43, fill=black!20
        ] (S44) {$X_4 \approx 18.0$};
      \node[draw,rounded rectangle,below = of S44, fill=black!20
        ] (S45) {$X_4 \approx 15.7$};
      \node[above=of S11] {Observe $Y_1$};
      \node[above=of S21] {Observe $Y_2$};
      \node[above=of S31] {Observe $Y_3$};
      \node[above=of S41] {Result \vphantom{$Y_1$}};

      \draw[-latex] (S12.east) -- (S21.west);
      \draw[-latex] (S12.east) -- (S23.west);
      \draw[-latex] (S13.east) -- (S22.west);
      \draw[-latex] (S13.east) -- (S24.west);
      \draw[-latex] (S13.east) -- (S25.west);

      \draw[-latex] (S22.east) -- (S32.west);
      \draw[-latex] (S22.east) -- (S33.west);
      \draw[-latex] (S22.east) -- (S31.west);
      \draw[-latex] (S24.east) -- (S34.west);
      \draw[-latex] (S24.east) -- (S35.west);

      \draw[-latex] (S31.east) -- (S43.west);
      \draw[-latex] (S32.east) -- (S44.west);
      \draw[-latex] (S33.east) -- (S41.west);
      \draw[-latex] (S34.east) -- (S45.west);
      \draw[-latex] (S35.east) -- (S42.west);

    \end{tikzpicture}
    \caption{%
      A resampling illustration for a bootstrap particle filter with 5
      simulations. The nodes are colored according to their weight---the darker
      nodes indicate more likely samples given the observation. The lines
      indicate how simulations survive, and possibly replicate, to the next
      step. No lines means a simulation is discarded. In the result, all
      samples of $X_4$ have the same weight, because there is no $Y_4$
      observation.
    }
  \label{fig:resample}
\end{figure}
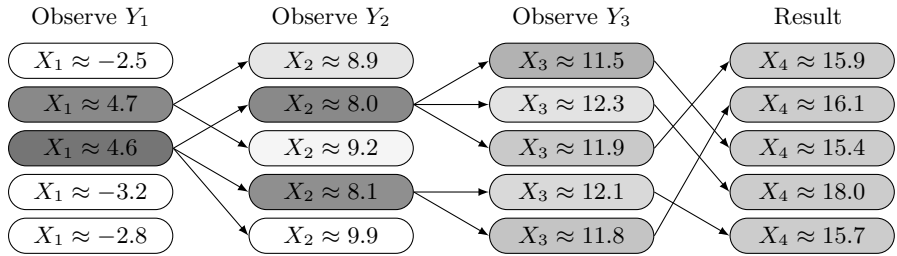

\paragraph{Sequential Monte Carlo.}
Consider again the example of a Bayesian network given in
Fig.~\ref{fig:baynet}. We are now interested in inferring the
\emph{marginal}\footnote{Meaning that we are only interested in some of the
unobserved random variables.} probability distribution $p(x_4 \mid
y_1,y_2,y_3)$---that is, the distribution over the next location of the moving
object given all of our observations up until this point. For this particularly
simple model, we can compute the exact solution in closed form by using
standard results from probability theory applied to the equations in
Fig.~\ref{fig:baynet}. In more complex
probabilistic models, an exact solution is most often not available. Instead,
approximate inference such as \gls{smc}~\cite{liu1998sequential} or
\gls{mcmc}~\cite{metropolis1953equation,hastings1970monte} methods must be
used. A basic Monte Carlo method is \emph{likelihood weighting}---simply
\emph{simulate} the model repeatedly, and weigh each simulation based on the
observed variables. This does not perform well for most models of interest, and
we can instead use an \gls{smc} method---the \emph{bootstrap
particle filter}~\cite{gordon2014probabilistic}.  The key idea in the bootstrap
particle filter is that we run many simulations in parallel, and
\emph{resample} simulations whenever encountering an observation.  Intuitively,
resampling means that less likely simulations are discarded and replaced by
more likely simulations.  This is illustrated in Fig.~\ref{fig:resample} for
the model in Fig.~\ref{fig:baynet}.  The resampling is especially obvious when
encountering the first observation $Y_1$---only two simulations of $X_1$ make
sense according to $Y_1$, and these simulations are the only two surviving to
the next step. In general, we can always run \gls{smc} inference on a Bayesian
network by finding a topological ordering over the random variables in the
network, and then simulating the network in that order. \Gls{smc} is, however,
not always the preferred method of inference, depending on the network
structure. \Gls{mcmc} is, for instance, sometimes a better alternative for
networks where observed nodes do not occur sequentially enough throughout the
network.

\begin{figure}[tb]
  \centering
  \begin{tikzpicture}[trim axis left, trim axis right]
    \begin{axis}[
      area style,
      yticklabel style={/pgf/number format/fixed},
      xlabel=$x_4$,
      ylabel={$p(x_4 \mid y_1,y_2,y_3)$},
      xmin=9,xmax=20,
      ymin=0,ymax=0.35,
      width=0.9\textwidth,
      height=4cm
      ]
      \addplot [ybar interval,mark=no,hist={density,bins=100}]
        table [y index=0] {case-study/histogram};

      \addplot [samples=100,dashed,thick,domain=9:20]
        {1 / sqrt(2*pi*1.6216216216216215) *
        exp(-(x-14.464864864864865)^2/(2*1.6216216216216215) )};
    \end{axis}
  \end{tikzpicture}
  \caption{%
    The result of running a bootstrap particle filter with $10\,000$ simulations
    for the model in Fig.~\ref{fig:baynet}. The normalized histogram shows the samples
    from the particle filter, and the dashed line shows the exact solution,
    which is available for this particular model.
  }
  \label{fig:hist}
\end{figure}
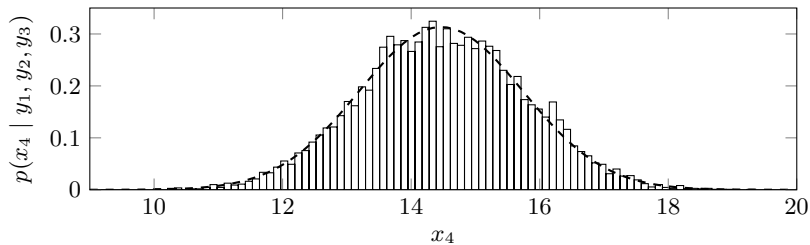

Fig.~\ref{fig:hist} shows a histogram of the samples produced by
running the bootstrap particle filter with $10\,000$ simulations (also
known as \emph{particles}) on the model in Fig.~\ref{fig:baynet}. Note
that the exact solution $p(x_4\mid y_1,y_2,y_3)$ is shown with the
dashed line. For more details on \gls{smc}, see
e.g., Doucet et al.  \cite{doucet2001introduction}.

\paragraph{Probabilistic programming: an example.}
\begin{figure}[tb]
  \centering
  \begin{tabular}{c}
    \begin{lstlisting}[basicstyle=\scriptsize\ttfamily,mathescape]
function sim(stop, lambda) {
  t = sample(exponential(lambda))
  if t <= stop then {
    weight(2.0)
    sim(stop-t, lambda+0.1)
  } else t
}

lambda = sample(gamma(1.0, 1.0))
stop = sample(gamma(1.0, 1.0))
sim = sim(stop, lambda)
weight(sim+lambda)
lambda
    \end{lstlisting}
  \end{tabular}
  \caption{A probabilistic program, written in our own functional, higher-order \gls{ppl}.}
  \label{fig:probprog}
\end{figure}
We gave a small toy example of a probabilistic program in
Section~\ref{sec:intro}. Here, we give a slightly bigger example, shown in
Fig.~\ref{fig:probprog}. The language contains a construct \texttt{sample} for
sampling from probability distributions, and a \texttt{weight} construct as seen before. The \texttt{sample} construct is equivalent to the unobserved random variables in a Bayesian network, and \texttt{weight} is related\footnote{%
  Observing a random variable $Y$ with probability distribution $p(y)$ as in a
  Bayesian network can be expressed as $\texttt{weight(}\log p(y)\texttt{)}$, where
  $y$ is the concrete observation.
}
to the observed random variables in a network.
The program is a smaller version of the
phylogenetic model used for the case study in Section~\ref{sec:case}, but still
demonstrates the alignment problem because \texttt{sim} recursively calls
itself from a stochastic branch (line 5) and contains a call to \texttt{weight}
(line 4). Hence, this call to \texttt{weight} should intuitively be marked
dynamic, since it might not be properly aligned.
Besides having stochastic
branches and recursion, probabilistic programming languages also differ from
Bayesian networks by defining an explicit ordering over random variables in the
program.  Such an ordering has to be provided separately for Bayesian networks
before performing inference.

\section{Discovering dynamic terms}\label{sec:disc}
As a first step, we perform a static analysis of our input program. The goal of
this analysis is to, for every term in the program, decide whether or not this
term can appear within a branch of an \texttt{if} expression with a stochastic
condition. We say that such a term is \emph{dynamic}. As we will see in
Section~\ref{sec:utilize}, the information produced by the analysis is key for
aligning the \gls{smc} inference correctly. We begin by introducing the target
language of the analysis. After this, we outline the analysis with examples and
give a formalization. Lastly, we discuss the limitations of the approach.

\subsection{The target language}
In order to simplify the presentation of the upcoming analysis, we begin by
introducing a \gls{ppl} with just enough constructs to make it universal.
Fig.~\ref{fig:lang} states the abstract syntax for such a language, based on
the untyped lambda calculus. Most importantly, the language contains \texttt{sample} and
\texttt{weight} constructs. Furthermore, the language also includes \texttt{if}
expressions, for which sampled values can be passed as the conditions. This,
together with the inherent recursion available in the untyped lambda calculus,
makes the language a minimal universal \gls{ppl}. Extending the language to a
more complete probabilistic programming language such as the language in
Fig.~\ref{fig:probprog} (which also contains various syntactic sugars) is
straightforward, and has been done for the case study in
Section~\ref{sec:case}.
\begin{figure}[tb]
  \[
    \begin{aligned}
      \expr \Coloneqq &\enspace \term^l\\
      \term \Coloneqq &\enspace
      x
      \enspace | \enspace
      c
      \enspace | \enspace
      \lambda x. \expr
      \enspace | \enspace
      \expr_1 \enspace \expr_2
      \enspace | \enspace
      \ttt{fix } \expr
      \enspace | \enspace
      \ttt{if } \expr_1 \ttt{ then } \expr_2 \ttt{ else } \expr_3
      \\
      | &\enspace
      \ttt{sample } \expr
      \enspace | \enspace
      \ttt{weight } \expr \\
      &\hspace{-8mm}
      \begin{aligned}
        x \in &\enspace X &
        &(\text{Variables})\\
        c \in &\enspace
        C &
        &(\text{Constants})\\
        %
        %
        %
        l \in &\enspace \mathbb{N} &
        &(\text{Labels}) \\
        &\hspace{-8mm}\{ \mathit{false}, \mathit{true}, () \}
          \cup \mathbb{R} \cup D \subseteq C
      \end{aligned}
    \end{aligned}
  \]
  \caption{A small \gls{ppl}. $D$ denotes a set of
  probability distributions, $()$ is the unit element.}
  \label{fig:lang}
\end{figure}
\begin{figure}[tb]
  \[
    \begin{gathered}
      \begin{aligned}
        \val \Coloneqq& \enspace
        c \enspace | \enspace
        \lambda x. \term \\
        \mathbf{F} \Coloneqq& \enspace
        \Box \enspace \term_2
        \enspace | \enspace
        \val_1 \enspace \Box
        \enspace | \enspace
        \ttt{fix } \Box
        \enspace | \enspace
        \ttt{if } \Box \ttt{ then } \term_2 \ttt{ else } \term_3
        \enspace | \enspace
        \ttt{sample } \Box
        \enspace | \enspace
        \ttt{weight } \Box \\[1em]
        &\hspace{-10mm}\boxed{\rightarrow}
      \end{aligned} \\
      \frac{\term \mid w \rightarrow \term' \mid w'}
      {F[\term] \mid w \rightarrow F[\term'] \mid w'}
      (\textsc{Cong}) \qquad
      \frac{}
      {(\lambda x. \term_1) \enspace \val_1 \mid w
      \rightarrow [x \mapsto \val_1]\term_1 \mid w}
      (\textsc{App}) \\
      \frac{}
      {\ttt{fix } (\lambda x. \term_1) \mid w \rightarrow
      [x \mapsto \ttt{fix } (\lambda x. \term_1)]\term_1 \mid w}
      (\textsc{Fix}) \\
      \frac{}
      {\ttt{if } \mathit{true} \ttt{ then } \term_2 \ttt{ else } \term_3
      \mid w \rightarrow \term_2 \mid w}
      (\textsc{IfTrue}) \\
      \frac{}
      {\ttt{if } \mathit{false} \ttt{ then } \term_2 \ttt{ else } \term_3
      \mid w \rightarrow \term_3 \mid w}
      (\textsc{IfFalse}) \\
      \frac{c \in D}
      {\ttt{sample} \enspace c \mid w \rightarrow \mathit{sample}(c) \mid w}
      (\textsc{Sample}) \qquad
      \frac{c \in \mathbb{R}}
      {\ttt{weight} \enspace c \mid w \rightarrow () \mid w + c}
      (\textsc{Weight}) \\
    \end{gathered}
  \]
  \caption{%
    An evaluation relation $\rightarrow$ for the language given in
    Fig.~\ref{fig:lang} with all labels ignored. The function $\mathit{sample}$
    correctly produces a sample from the provided distribution.  All congruence
    rules are compactly described by \textsc{Cong}, which specifies one rule
    for every case in $\mathbf{F}$. $\mathbf{F}[\term]$ means that we replace
    the $\Box$ in one case in $F$ with $\term$.
  }
  \label{fig:sem}
\end{figure}

For convenience when later defining our algorithm, we split the language into
two production rules, $\expr$ and $\term$, where $\expr$ is a \emph{labeled}
version of $\term$. For all programs in the language, we assume a unique
labeling of all expressions, and that all variables are bound in at most one
place (which means that all variable names are unique). Any program can be
transformed to fulfill this without any input from the programmer. The unique
labels and variables are requirements for the static analysis.

Also included in the language is a set of constants $C$. We leave this set
unspecified, with booleans, real numbers, and the unit element as exceptions.
The reason for explicitly including booleans in the set of constants is because
they are needed for \texttt{if} expressions.  Additionally, real numbers are
needed as arguments for \texttt{weight}, and the unit element as the result of
a call to \texttt{weight}. We also assume that various probability
distributions $\mathit{dist} \in D$ from which to sample are included in the set
of constants, $C$. We do, however, limit these distributions to not range over
lambda abstractions, since this would complicate the analysis significantly.

Lastly, the language includes an explicit fixpoint operator \texttt{fix}. Since
we are dealing with the untyped lambda calculus, we could construct such an
operator (the $Y$ combinator) in the language itself. There is, however, an
important difference between the two: the explicit fix point operator cannot be
passed around as a value---it must be applied directly. As a consequence, we
can make the analysis less conservative. That is, fewer terms will be marked as
dynamic in comparison to using the $Y$ combinator.

To give some more intuition for the language, we give a small-step operational
semantics for it in Fig.~\ref{fig:sem}. It is an ordinary call-by-value
semantics for the untyped lambda calculus, with a weight $w$ added in the
evaluation relation. This weight is updated at calls to \texttt{weight}, which
is reflected in the rule \textsc{Weight}. This semantics corresponds to
obtaining a single sample from the distribution encoded by the program in a
likelihood weighting inference algorithm.  Likelihood weighting was briefly
mentioned in Section~\ref{sec:prelim}. We will see how this semantics relates
to \gls{smc} and resampling in Section~\ref{sec:utilize}.

\subsection{The analysis}
Finding dynamic terms is not straightforward, as can be seen from two simple examples. The
first example is given by the following program (labels omitted)
\begin{equation}\label{eq:ex1}
  (\lambda x. \ttt{if } \ttt{sample} \enspace \mathit{dist} \ttt{ then }
  (x \enspace c_1)
  \ttt{ else } c_2) \enspace (\lambda y. y),
\end{equation}
where $\mathit{dist}$ is a distribution over booleans and $c_1$ and $c_2$ are constants.
The analysis result for this program should, intuitively, be
\begin{equation}\label{eq:ex1s}
  (\lambda x. \ttt{if } \ttt{sample} \enspace \mathit{dist} \ttt{ then }
  \underline{(\underline{x} \enspace \underline{c_1})}
  \ttt{ else } \underline{c_2})
  \enspace \underline{(\lambda y. \underline{y})}
\end{equation}
where the underlining shows all parts of the program which can appear within a
stochastic branch. The right-hand side of the outermost application is bound by
the left-hand side lambda abstraction, and can therefore appear in one of the
branches. By regarding the entire program as a tree structure, we see that
information has been propagated from the left-hand side of a node, to its
right-hand side.

The reverse is also possible. Consider the following program:
\begin{equation}\label{eq:ex2}
    (\lambda a. (\lambda b. a \enspace b) \enspace
    (\lambda c. c)) \enspace
  (\lambda d. \ttt{if } \ttt{sample } \mathit{dist} \ttt{ then }
    (d \enspace c_1)
    \ttt{ else } c_2)
\end{equation}
The analysis result for this program is given by
\begin{equation}
    (\lambda a. (\lambda b. a \enspace b) \enspace
    \underline{(\lambda c. \underline{c})}) \enspace
    (\lambda d. \ttt{if } \ttt{sample } \mathit{dist} \ttt{ then }
    \underline{(\underline{d} \enspace \underline{c_1})}
    \ttt{ else } \underline{c_2}),
\end{equation}
showing that information from the right-hand side of a node can propagate to
its left-side.

We propose a solution for finding dynamic terms based on \emph{0-CFA}, a control-flow analysis
algorithm for higher-order functional programming languages originally
introduced by Shivers \cite{shivers1988control,shivers1991control}. The 0 in
0-CFA stands for \emph{context insensitivity}. Many other, less conservative,
approaches to control-flow in higher-order functional languages also exist
\cite{midtgaard2012control}. We give details on the limitations of context insensitivity in Section~\ref{sec:limitations}   An example of a more accurate analysis is $k$-CFA,
where $k$ levels of context sensitivity are included in the analysis. This
causes the analysis to run in exponential time, already for $k = 1$. 0-CFA has
worst-case time complexity $O(n^3)$, where $n$ is the size of the program. This
is an upper bound, and might not affect how large programs can be handled in
practice.
The version of 0-CFA that we present here is based on Nielson et al.
\cite{nielson1999principles}.

\paragraph{Generating the constraints.}
To give some intuition for the algorithm, we describe it with the program
\eqref{eq:ex1} as a running example.
The first step is to assign each subterm a unique label:
\begin{equation}\label{eq:exlabel}
  ((\lambda x. (\ttt{if } (\ttt{sample} \enspace \mathit{dist}^1)^2 \ttt{ then }
  (x^3 \enspace c_1^4)^5
  \ttt{ else } c_2^6)^7)^8 \enspace (\lambda y. y^9)^{10})^{11}
\end{equation}
As we will see, this labeling enables reasoning about possible flows of control
in the program.  We also define
$ \mathbf{T} = \{ \enspace (\lambda x. \cdot^7)^8, (\lambda y. \cdot^9)^{10} \enspace \}$,
which is the set of all lambda terms in the program. The bodies of the lambda
terms are replaced by $\cdot$, since they are not required in the analysis.
Next, we generate a set of \emph{constraints} for the program. These
constraints capture how both data and lambdas might flow between different
locations in the program. Our goal is to find a minimal assignment to the
\emph{unknown sets} occurring in the constraints, such that the constraints are
not violated. Such a solution is guaranteed to exist, and is key to finding
all dynamic terms. The constraints generated for \eqref{eq:exlabel} are
\begin{equation}\label{eq:genex}
  \begin{aligned}
    \mathit{gen}(t) = \{ \enspace
      &\{ \mathbf{stoch} \} \subseteq S_{2},
      \{ (\lambda y. \cdot^9)^{10} \} \subseteq S_{10},
      \{ (\lambda x. \cdot^7)^8 \} \subseteq S_{8}, \\
      &S_{y} \subseteq S_{9},
      S_{5} \subseteq S_{7},
      S_{6} \subseteq S_{7},
      S_{x} \subseteq S_{3}, \\
      &\{ (\lambda x. \cdot^7)^8 \} \subseteq S_{8}
      \Rightarrow S_{10} \subseteq S_{x},
      \{ (\lambda x. \cdot^7)^8 \} \subseteq S_{8}
      \Rightarrow S_{7} \subseteq S_{11},\\
      &\{ (\lambda y. \cdot^9)^{10} \} \subseteq S_{8}
      \Rightarrow S_{10} \subseteq S_{x},
      \{ (\lambda y. \cdot^9)^{10} \} \subseteq S_{8}
      \Rightarrow S_{9} \subseteq S_{11},\\
      &\{ (\lambda x. \cdot^7)^8 \} \subseteq S_{3}
      \Rightarrow S_{4} \subseteq S_{x},
      \{ (\lambda x. \cdot^7)^8 \} \subseteq S_{3}
      \Rightarrow S_{7} \subseteq S_{5},\\
      &\{ (\lambda y. \cdot^9)^{10} \} \subseteq S_{3}
      \Rightarrow S_{4} \subseteq S_{y},
      \{ (\lambda y. \cdot^9)^{10} \} \subseteq S_{3}
      \Rightarrow S_{9} \subseteq S_{5}
    \enspace \}
  \end{aligned}
\end{equation}
The variables $S_1,S_2,\ldots,S_{11}, S_x, S_y$ denotes the unknown sets
associated with each label or variable in the program.  There are three types
of constraints: direct, flow, and implication flow
constraints.  \emph{Direct} constraints force a set $S$ to contain a single
\emph{abstract value} $\aval$, which can either be $\mathbf{stoch}$ or a lambda
abstraction:
$\aval \Coloneqq \enspace \mathbf{stoch} \enspace
| \enspace (\lambda x. \cdot^{l_1})^{l}$.
The first constraint in \eqref{eq:genex}, $\{ \mathbf{stoch} \}
\subseteq S_{2}$, states that the term at label 2 in the program may be stochastic.
By looking at \eqref{eq:exlabel}, this is clearly true---the term at label 2
contains a sample from a distribution. We also have two other direct
constraints, which states that lambda expressions may occur at the label where
they syntactically originate. This must also clearly be true. The flow and
implication flow constraints state how the abstract values flow between the
sets.  \emph{Flow} constraints declare an immediate link between two sets. For
instance, two of the flow constraints state that $S_5$ and $S_6$ must flow to
$S_7$, because the \texttt{if} expression at label 7 can evaluate to both its branches.
\emph{Implication flow} constraints, on the other hand, states that if an abstract
value is in one set, this causes a flow between other sets. One such constraint
is $\{ (\lambda y.
\cdot^9)^{10} \} \subseteq S_{3} \Rightarrow S_{4} \subseteq S_{y}$ which
states that if the lambda with variable $y$ occurs at the term with label 3,
then the term at label 4 must flow to the variable y. This is a simple
consequence of how applications are evaluated. Formally, the constraints are given by
\begin{equation}
  \begin{aligned}
    \set  \Coloneqq &\enspace S_l \enspace | \enspace S_x \\
    \cstr \Coloneqq &\enspace \{ \aval \} \subseteq \set &
    &(\text{Direct})\\
    | &\enspace \set_1 \subseteq \set_2 &
    & (\text{Flow})\\
    | &\enspace \{ \aval \} \subseteq \set_1 \Rightarrow \set_2 \subseteq \set_3 &
    & (\text{Implication flow})
  \end{aligned}
\end{equation}

\begin{figure}[tb]
  \[
    \begin{aligned}
      &\mathit{gen}(x^l) = \{S_{x} \subseteq S_{l}\} \\
      &\mathit{gen}(c^l) = \varnothing  \\
      &
      \begin{aligned}
        \mathit{gen}((\lambda x. \term^{l_1})^{l})
        = \enspace
        &\{\{(\lambda x. \term^{l_1})^{l}\} \subseteq S_{l} \}
        \cup \mathit{gen}(\term^{l_1})
      \end{aligned} \\
      &
      \begin{aligned}
        \mathit{gen}((\term_1^{l_1} \enspace \term_2^{l_2})^{l})
        = \enspace \mathit{gen}(\term_1^{l_1}) \cup \mathit{gen}(\term_2^{l_2})
        &\cup \{
          \{\term\} \subseteq S_{l_1} \Rightarrow S_{l_2} \subseteq S_{x}
          \mid \term =
          (\lambda x. \term_3^{l_3})^{l_4} \in \mathbf{T}
        \} \\
        &\cup \{
          \{\term\} \subseteq S_{l_1} \Rightarrow S_{l_3} \subseteq S_{l}
          \mid \term =
          (\lambda x. \term_3^{l_3})^{l_4} \in \mathbf{T}
        \}
      \end{aligned} \\
      &
      \begin{aligned}
        \mathit{gen}((\ttt{fix } \term^{l_1})^{l})
        = \enspace \mathit{gen}(\term^{l_1})
        &\cup \{
          \{\term\} \subseteq S_{l_1} \Rightarrow S_{l_2} \subseteq S_{x}
          \mid \term =
          (\lambda x. \term^{l_2})^{l_3} \in \mathbf{T}
        \} \\
        &\cup \{
          \{\term\} \subseteq S_{l_1} \Rightarrow S_{l_2} \subseteq S_{l}
          \mid \term =
          (\lambda x. \term^{l_2})^{l_3} \in \mathbf{T}
        \} \\
      \end{aligned} \\
      &
      \begin{aligned}
        \mathit{gen}((\ttt{if } \term_1^{l_1} \ttt{ then }
        \term_2^{l_2} \ttt{ else } \term_3^{l_3})^l) =
        &\enspace \mathit{gen}(\term_1^{l_1}) \cup \mathit{gen}(\term_2^{l_2}) \cup \mathit{gen}(\term_3^{l_3}) \\
        &\enspace \qquad \cup \{S_{l_2} \subseteq S_{l}\} \cup \{S_{l_3} \subseteq S_{l}\}
      \end{aligned} \\
      &
      \begin{aligned}
        \mathit{gen}((\ttt{sample } \term^{l_1})^l) = \mathit{gen}(\term^{l_1}) \cup
        \{\{\mathbf{stoch}\} \subseteq S_{l}\} \\
      \end{aligned} \\
      &
      \begin{aligned}
        \mathit{gen}((\ttt{weight } \term^{l_1})^l) = \mathit{gen}(\term^{l_1})
      \end{aligned} \\
    \end{aligned}
  \]
  \caption{The constraint generation function $\mathit{gen}$.}
  \label{fig:gen}
\end{figure}
The constraint generation function $\mathit{gen}$ is defined recursively in
Fig.~\ref{fig:gen}. The most intricate part of $\mathit{gen}$ is the constraint
generation for applications and fixpoints. Both produce two flow implication
constraints for each lambda in $\mathbf{T}$, which we defined earlier. The
application case is fairly intuitive: if a lambda can occur at the left hand
side of an application, it must be the case that the right hand side flows to
the variable bound by the lambda, and that the term enclosed in the lambda can
flow to the result of the application. Fixpoints are a bit more difficult. If a
lambda term $(\lambda x. \term^{l_2})^{l_3}$ can occur as the argument to a
\texttt{fix} operator, two things must hold. Because of how \texttt{fix} is
defined, the enclosed lambda term with label $l_2$ is the actual (recursive)
function being computed---\texttt{fix} simply binds the function itself to the
variable $x$. Therefore, label $l_2$ must flow to $x$ since we need to be able
to use the function recursively through this binding, and $l_2$ can also flow
to $l$, because $l_2$ is the actual function produced by the \texttt{fix}
operator.

\paragraph{Solving the constraints.}
In order to solve the constraints, we refer to the full description of 0-CFA in
Nielson et al. \cite{nielson1999principles}. For the constraints in
\eqref{eq:genex}, the minimal solution is given by
\begin{equation}\label{eq:gensol}
  \begin{aligned}
    &S_y  = \varnothing &
    &S_x  = \{ (\lambda y. \cdot^9)^{10} \} &
    &S_1  = \varnothing &
    &S_2  = \{ \mathbf{stoch} \} \\
    &S_3  = \{ (\lambda y. \cdot^9)^{10} \} &
    &S_4  = \varnothing &
    &S_5  = \varnothing &
    &S_6  = \varnothing \\
    &S_7  = \varnothing &
    &S_8  = \{ (\lambda x. \cdot^7)^8 \} &
    &S_9  = \varnothing &
    &S_{10}  = \{ (\lambda y. \cdot^9)^{10} \} \\
    &S_{11}  = \varnothing.
  \end{aligned}
\end{equation}
This can easily be verified to be a minimal solution satisfying all the
constraints in \eqref{eq:genex}.

\paragraph{Finding the dynamic terms.}
\begin{algorithm}[tb]
  \caption{%
    The final phase of the analysis. Uses the 0-CFA output to discover dynamic
    parts of the program. The input consists of the labeled program $\term^l$,
    and the results of the 0-CFA analysis $S$ (that is, all the sets produced
    by the analysis). The function $\mathit{labels}$ returns all labels within
    a term. The function $\mathit{subexpr}$ returns all direct subexpressions
    of a term $\term$.
}
  \label{alg:disc}
  \begin{algorithmic}[1]
    \Function{Dynamic}{$\term^l$, $S$}
    \For{$l' \in \mathit{labels}(\term^l)$} $\mathit{Dyn}(l') \gets \mathit{false}$
    \Comment{Initialization} \EndFor
    \State $mod \gets \mathit{true}$
    \While{$mod$} \Comment{Iterate until fixpoint}
    \State $mod \gets \mathit{false}$; \Call{Recurse}{$\mathit{false}$, $\term^l$}
    \EndWhile
    \State \Return $\{ l \mid l \in labels(\term^l), \mathit{Dyn}(l) = \mathit{true} \}$
    \EndFunction
    \State

    \Function{Recurse}{$\mathit{flag}$, $\term^l$}
    \If{$\mathit{flag} \lor \mathit{Dyn}(l)$} \Comment{Mark dynamic terms}
    \If{$\neg \mathit{Dyn}(l)$}
    \State $\mathit{Dyn}(l) \gets \mathit{true}$
    \State $mod \gets \mathit{true}$
    \EndIf
    \For{$(\lambda x. \cdot^{l_1})^{l_2} \in S_l$}
    \If{$\neg \mathit{Dyn}(l_2)$}
    \State $\mathit{Dyn}(l_2) \gets \mathit{true}$
    \State $mod \gets \mathit{true}$ \EndIf
    \EndFor
    \EndIf
    \State \textbf{match} $t$ \textbf{with}
    \State \hspace{3mm}
    $\ttt{if } \term_1^{l_1} \ttt{ then }
    \term_2^{l_2} \ttt{ else } \term_3^{l_3}$:
    \Comment{Detect stochastic branches}
    \State \hspace{6mm} \Call{Recurse}{$\mathit{flag}$,$\term_1^{l_1}$}
    \State \hspace{6mm} $\mathit{flag} \gets \mathit{flag} \lor \mathbf{stoch} \in S_{l_1}$
    \State \hspace{6mm} \Call{Recurse}{$\mathit{flag}$,$\term_2^{l_2}$};
    \Call{Recurse}{$\mathit{flag}$,$\term_3^{l_3}$}
    \State \hspace{3mm}
    $\lambda x. \term_1^{l_1}$:
    \Comment{Detect previously marked lambdas}
    \State \hspace{6mm} \Call{Recurse}{$\mathit{Dyn}(l) \lor \mathit{flag}$, $\term_1^{l_1}$}
    \State \hspace{3mm} otherwise:
    \textbf{for} $\term_1^{l_1} \in \mathit{subexpr}(\term)$ \textbf{do}
    \Call{Recurse}{$\mathit{flag}$, $\term_1^{l_1}$}
    \EndFunction
  \end{algorithmic}
\end{algorithm}
The last step is to use the 0-CFA results to find dynamic terms. To do this, we do a
depth-first left-to-right traversal of the program, flagging all terms (or,
equivalently, their labels) occurring in the branch of a stochastic branch as
dynamic. We can identify stochastic branches by checking if $\mathbf{stoch}$ is
a member of $S_l$, where $l$ is the label of the condition term of an \texttt{if}
expression. In \eqref{eq:exlabel}, during traversal, we first go down the left
branch of the outermost application and eventually reach
\begin{equation}
(\ttt{if } (\ttt{sample} \enspace \mathit{dist}^1)^2 \ttt{ then }
  (x^3 \enspace c_1^4)^5
  \ttt{ else } c_2^6)^7.
\end{equation}
We see that $\mathbf{stoch}$ is in $S_2$, and the branch is therefore
stochastic and we recursively flag the terms in the branches. Additionally, we
flag the lambda term $(\lambda y. y^9)^{10}$, since it is in the set $S_3$.
Because of this, when we return to the outermost application and traverse down
the right hand side, we can see that $(\lambda y. y^9)^{10}$ is flagged.
Therefore, we also flag all terms enclosed in this lambda, which in this case
is $y^9$. To summarize, the result of performing this analysis on
\eqref{eq:exlabel} with the help of \eqref{eq:gensol} is
$\{ 3, 4, 5, 6, 9, 10 \}$,
which matches the result in \eqref{eq:ex1s} with the labels in
\eqref{eq:exlabel}.  Note that $y^9$ would not have been flagged if we would
have done a right-to-left traversal. In general, we need to repeatedly traverse
the program until fixpoint, allowing all terms reachable from a stochastic
branch to be flagged as dynamic. The complete algorithm is shown in
Algorithm~\ref{alg:disc}. We can reason about the time complexity as follows:
on every iteration, at least one label is flagged or the program terminates.
Since we have $n$ labels, where $n$ is the size of the program, and every
iteration is performed in $n$ steps, it follows that the algorithm (in the
worst case) terminates in $O(n^2)$ steps---less than the $O(n^3)$ for the 0-CFA
analysis. Therefore, the overall complexity is still $O(n^3)$.

\subsection{Limitations}\label{sec:limitations}
The main limitation of the algorithm presented in this paper is the lack of
context sensitivity in the analysis. In practice, this will cause problems when
reusing functions in both stochastic and non-stochastic contexts---the
non-stochastic contexts will sometimes be unnecessarily marked as stochastic.
As an example, consider running the analysis on a program written in the same
language as in Fig.~\ref{fig:introex} and Fig.~\ref{fig:probprog}:
\\[1em]
\texttt{function plus(a, b) \string{ a + b \string}} \\
\texttt{plus(sample(normal(0,1)), 2)} \\
\texttt{if plus(1, 3) < 5 then \underline{true} else \underline{false}}
\\[1em]
Our analysis has marked the branches of the \texttt{if} expression as
dynamic, even though the condition is clearly not stochastic.  This is because
of context insensitivity: the analysis cannot distinguish between the two
applications of plus. Since one of the applications produces a stochastic
value, \emph{all} applications of plus in the program are marked as
stochastic---even if they are in fact not stochastic. In this paper, we avoid
this problem by using built in operators which cannot be passed around the
program as values in the same way as user-defined lambda abstractions. This
makes the analysis less conservative when using 0-CFA. An obvious direction for
future work is exploring other approaches to higher-order control flow analysis
that do take context into account \cite{midtgaard2012control}.

\section{Utilizing the Analysis Results for Sequential Monte Carlo Inference}
\label{sec:utilize}
In this section, we use the analysis result from Section~\ref{sec:disc} to
transform the input program, enabling \emph{aligned} \gls{smc} inference. Most
importantly, we indicate how to modify the semantics of Fig.~\ref{fig:sem} to
accommodate such inference, and also give the aligned \gls{smc} algorithm for
probabilistic programming. We use the program from Section~\ref{sec:prelim},
Fig.~\ref{fig:probprog} as a running example, assuming that the semantics
include proper extensions for arithmetic and comparison.

\paragraph{Transforming the program.}
We begin by extending our language with one additional construct:
\texttt{dweight} (dynamic weight). In contrast to \texttt{weight}, the
\texttt{dweight} construct will not cause resampling to be
performed. By using the information about dynamic terms from our static
analysis, we do a simple transformation of our input program: we replace all
dynamic \texttt{weight} terms with \texttt{dweight} (ignoring
labels, since they are no longer required).  The remaining calls to
\texttt{weight} are now \emph{aligned}---they are (1) always executed, and (2) always
executed in the same order. This is a simple consequence of them not being
reachable from stochastic branches.  The transformation allows \gls{smc}
inference to only resample at aligned calls to \texttt{weight} in the original
program.  As an example, for the program in Fig.~\ref{fig:probprog}, the call
to \texttt{weight} at line 4 will be replaced by \texttt{dweight}, and the
\texttt{weight} at line 13 will be untouched.

\paragraph{Modifying the semantics.}
Next, we modify our semantics to support \gls{smc} inference. In
order to do this, we first need to do another program transformation to enable
\emph{pausing} and \emph{resuming} executions when resampling. We will not go
into detail about this transformation here, but the result is a program in
\emph{\gls{cps}} \cite{steele1978rabbit,appel2007compiling}. Such a
transformation is commonly used in \glspl{ppl}, for instance in WebPPL and
Anglican.  The essential property of having the program in \gls{cps} is that
functions never return. Instead, every function takes an additional argument, a
\emph{continuation} function, which is applied to the result of the function
application in order to continue evaluation. The continuation can be thought of
as a representation of the call stack that is explicitly available at each
function call. In essence, this
enables us to modify our evaluation relation $\rightarrow$ so that we can pause
and resume evaluation at calls to \texttt{weight}. To enable pausing, we
explicitly add a \texttt{pause} term in our language. In the
\gls{cps} transformed language, the \texttt{weight}, \texttt{dweight}, and
\texttt{pause} terms all take one extra continuation argument $\term_c$. That
is, $ \term \Coloneqq \enspace \ldots \enspace | \enspace
\texttt{weight } \term_c \enspace \term \enspace | \enspace
\texttt{dweight } \term_c \enspace \term \enspace | \enspace \texttt{pause } \term_c$.
The key modification in the semantics for \texttt{weight} is shown in
Fig.~\ref{fig:modsem}. For \texttt{dweight}, we simply update the weight and take a step to $\term_1$, the body of the continuation.  This is a \gls{cps}
equivalent of the previous rule for \texttt{weight} in Fig.~\ref{fig:sem}. For
the new \texttt{weight}, we instead want to indicate to the inference algorithm
that the program is paused.  Therefore, we return a \texttt{pause} term, with
the continuation as argument. There is no evaluation rule for \texttt{pause},
so the evaluation halts, and it is up to the \gls{smc} inference algorithm to
decide the next course of action.
\begin{figure}[tb]
  \[
    \begin{gathered}
      \frac{c \in \mathbb{R}}
      {\ttt{dweight} \enspace (\lambda x. \term_1) \enspace c \mid w \rightarrow
      \term_1 \mid w + c}
      (\textsc{DWeightCPS}) \\
      \frac{c \in \mathbb{R}}
      {\ttt{weight} \enspace (\lambda x. \term_1) \enspace c \mid w \rightarrow
      \texttt{pause } (\lambda x. \term_1) \mid w + c}
      (\textsc{WeightCPS})
    \end{gathered}
  \]
  \caption{The \gls{cps} evaluation rules for \texttt{dweight} and
  \texttt{weight}}
  \label{fig:modsem}
\end{figure}

\paragraph{Aligned sequential Monte Carlo.}
\begin{algorithm}[tb]
  \caption{%
    The algorithm for aligned \gls{smc} inference in probabilistic programs.
    The $\mathit{eval}$ function repeatedly applies $\rightarrow$ on a program
    $t$ with weight $w$ until no evaluation rule is applicable. The input
    $n$ gives the number of executions, or particles.
  }\label{alg:smcalign}
  \begin{algorithmic}[1]
    \Function{AlignedSMC}{$t$, $n$}
    \For{$i \gets 1$ to $n$}
    $t_i \gets t$ \Comment{Create $n$ copies of $t$}
    \EndFor
    \For{$i \gets 1$ to $n$}
    $r_i \gets \mathit{eval}(t_i,0)$
    \EndFor
    \While{$r_1 = \texttt{pause } (\lambda x. \term) \mid w_1$}
    \Comment{Check if \texttt{weight} has been encountered.}
    \For{$i \gets 1$ to $n$}
    $(\texttt{pause }t_i \mid w_i) \gets r_i$
    \EndFor
    \State $t_{1:n} \gets \mathit{resample}(t_{1:n}, w_{1:n})$
    \For{$i \gets 1$ to $n$}
    $r_i \gets \mathit{eval}(t_i \, (),0)$
    \EndFor
    \EndWhile
    \For{$i \gets 1$ to $n$}
    $(t_i \mid w_i) \gets r_i$
    \EndFor
    \State $t_{1:n} \gets \mathit{resample}(t_{1:n}, w_{1:n})$
    \State \Return $\{ t_1, t_2, \ldots, t_n \}$
    \EndFunction
  \end{algorithmic}
\end{algorithm}
The algorithm for aligned \gls{smc} is shown in Algorithm~\ref{alg:smcalign}.
The intuition is quite simple: do $n$ executions of the program using
$\rightarrow$ and stop whenever encountering an aligned \texttt{weight} to
resample before continuing the executions by applying $()$ to the
continuations. Note that we use the alignment property at line 4, assuming that
if $r_1$ is a \texttt{pause} term, then this will also be true for all other
$r_i$. Also note that we set all weights to 0 after resampling. This is because
resampling, by definition, produces a set of unweighted samples (in our case
executions) from a set of weighted samples.
When finished, the $eval$ function will return a final value
with an attached weight. After doing a final resample (the weights can have
been modified by calls to \texttt{dweight} since the last resample), the values
are returned as samples. For the program in Fig.~\ref{fig:probprog}, the
algorithm would run all particles until encountering the single call to
\texttt{weight} (line 13), accumulating the weights for each particle when
encountering differing number of calls to \texttt{dweight} (line 6).  Hence,
there will only be two resamples: one at the \texttt{weight} at line 13, and
one at the end of the program.

\section{Case study}\label{sec:case}
In this section, we give the details on a case study for a probabilistic model
from phylogenetics, expressed as a probabilistic program. We begin by briefly
describing the implementation of the analysis presented in
Sections~\ref{sec:disc}~and~\ref{sec:utilize}. This is followed by a
description of the model, and the quantity of interest that we wish to estimate
using \gls{smc}. Lastly, we present the results of the case study in the form
of a comparison between aligned and unaligned \gls{smc}, and discuss the main
limitations of our algorithms. All source code used in this case study is
available at \url{https://github.com/miking-lang/pplcore}.

\paragraph{Implementation.}
The implementation language extends the abstract syntax and semantics of the
language in Fig.~\ref{fig:lang} and Fig.~\ref{fig:sem} with various operators
for arithmetic and comparison. Examples of the concrete syntax is given in
Fig.~\ref{fig:introex} and Fig.~\ref{fig:probprog}. Our implementation of
aligned \gls{smc} implements the analysis from Section~\ref{sec:disc}, and
follows Algorithm~\ref{alg:smcalign}.  Additionally, we implement unaligned
\gls{smc}, based on the approach used in WebPPL~\cite{dippl}. In this version
of \gls{smc}, dynamic calls to weight also participate in resampling, as well
as executions that have already terminated (which can occur since alignment is
not guaranteed).  We use \emph{systematic resampling} \cite{douc2005comparison}
for both versions of \gls{smc}. Everything is implemented in OCaml.

\paragraph{The model and the inference problem.}
We test the performance of the algorithms by using an example from statistical
phylogenetics, in which a birth-death model is used to describe the rates of
speciation and extinction in a group of organisms. Such models are of
considerable interest to evolutionary biologists, as they can be used to study
many important phenomena, such as the effects of various life-history traits or
of environmental factors on net diversification rates~\cite{nee2006birth}. A famous
research problem that can be addressed using birth-death models is the question
of whether the extinction of dinosaurs at the end of the Cretaceous epoch caused
an increased diversification rate in mammals~\cite{ronquist2016closing}.

In typical cases, we only have reliable observations of the extant species of
the group, that is, the lineages that have survived until the present---the
extinct lineages are unknown to us. From DNA sequence data and calibration
fossils, we can reconstruct a time tree that describes how and when the extant
lineages diverged from each other; this is known as a \emph{reconstructed
tree}~\cite{nee1994reconstructed}. The task is now to estimate the speciation
(birth) and extinction (death) rates from such a reconstructed time tree.

We focus on the basic task of estimating the \emph{normalizing constant} for a
particular set of birth and death rates and a given reconstructed time tree.
That is, we force the model to always produce the same sample of the rates, and
instead produce estimates of how likely this sample is given the data.  The
logarithm of this quantity can be estimated with \gls{smc} through
\begin{equation}\label{eq:norm}
  \sum_{t=1}^T\left(
  \log \sum_{i=1}^N \exp(w_t^i) - \log N
  \right)
\end{equation}
where $t$ ranges over all resampling points in the program, and
$w_i^t$ denotes the weight of execution $i$ at resampling point $t$. $N$ is the
total number of executions.
The normalizing constant can be used for Bayesian model comparison of different
scenarios; they can also be used in a nested particle \gls{mcmc}
approach, in which \gls{smc} is combined with \gls{mcmc} to estimate a
posterior distribution over birth and death rates.

Specifically, we use a consensus estimate of the divergence times of the
28 extant species of pitheciid monkeys provided by the TimeTree
project~\cite{timetree}. The tree has one trichotomy involving
\emph{Chiropotes albinasus}. We resolve this ambiguity by assuming that
\emph{C. albinasus} belongs to \emph{Chiropotes}, and that the stem lineage of
\emph{Chiropotes} existed for 0.2 Ma before branching into extant species. This
is similar to the shortest branch length observed in other parts of the tree.
The birth rate is set to 0.2 Ma-1 and the death rate to 0.1 Ma-1.

In summary, the input data is a tree over which we simulate a birth-death
process.  Because of this, we have a mix of aligned and unaligned calls to
\texttt{weight}---aligned calls occur when traversing the nodes in the input
tree, and unaligned calls occur when simulating along edges in the tree.

\begin{figure}[tb]
  \centering
  \begin{tikzpicture}
    \begin{axis}[trim axis left, trim axis right, width=0.77\textwidth,
      height=5cm,
      ymin=0,ymax=7,
      ytick={1,2,3,4,5,6},
      yticklabel style={align=right},
      yticklabels={%
        {Aligned, $10\,000$ executions},
        {Aligned, $1000$ executions},
        {Aligned, $100$ executions},
        {Unaligned, $10\,000$ executions},
        {Unaligned, $1000$ executions},
        {Unaligned, $100$ executions},
      }]
      \addplot [boxplot]
        table [y index=0] {case-study/smca10000};
      \addplot [boxplot]
        table [y index=0] {case-study/smca1000};
      \addplot [boxplot]
        table [y index=0] {case-study/smca100};

      \addplot [boxplot]
        table [y index=0] {case-study/smcu10000};
      \addplot [boxplot]
        table [y index=0] {case-study/smcu1000};
      \addplot [boxplot]
        table [y index=0] {case-study/smcu100};

      \addplot [mark=none,dashed] coordinates{%
          (-56.33242285520951, 0)
        (-56.33242285520951, 7)};
    \end{axis}
  \end{tikzpicture}
  \caption{The result of the case study, showing the increase in accuracy from
  using aligned \gls{smc}.}
  \label{fig:result}
\end{figure}
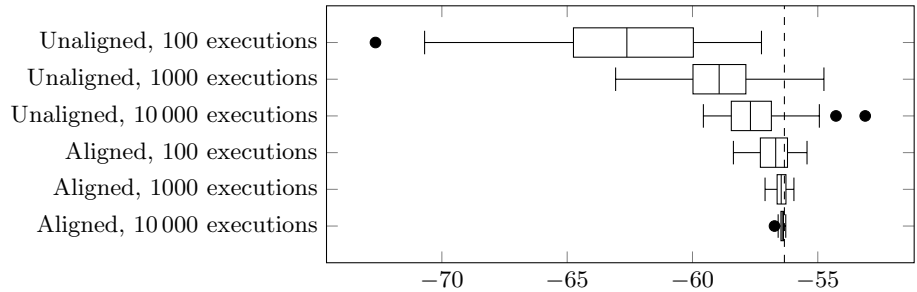

\paragraph{Result.}
The result of our case study is shown in Fig.~\ref{fig:result}, using box plots
for $100$ estimates produced from \eqref{eq:norm} on our phylogenetic model for
different number of executions. The exact solution is available analytically
for this model and is shown with the dashed line. We see that the aligned
version gives better estimates in all cases. In addition, we measured aligned
\gls{smc} to be approximately 1.66 times faster on average than unaligned
\gls{smc} for this model.

\paragraph{Discussion.}
Looking back at the nonoptimal results for unaligned \gls{smc} in
Section~\ref{sec:intro}, the improvement of aligned over unaligned \gls{smc}
(for the same number of executions) in this case study intuitively makes sense.
However, it seems that, even when using the unaligned version, the result
converges on the true value as the number of total executions increase. We can
make the same observation for the example in Fig.~\ref{sec:intro}, if we reduce
the differences between the \texttt{weight}s. By, for instance, setting
\texttt{weight(85)} to \texttt{weight(5)}, and \texttt{weight(95)} to
\texttt{weight(15)}, we do get approximately the same number of executions
(taking the weights into account) for
each branch when running enough executions in total. As
long as a single execution from the \texttt{false} branch survives, it will
have much higher
weight in the end, thus offsetting the bias in the initial resampling. This
implies that unaligned \gls{smc} is most likely correct, but that an enourmous
number of executions might be required, even for very simple models such as the
model in Fig.~\ref{fig:introex}.

The increase in speed from using aligned \gls{smc} most likely comes from
simply doing less resampling while running the \gls{smc} algorithm.

\section{Related work}

Naturally, the work most closely related to ours can be found in papers on
universal probabilistic programming languages using \gls{smc}, such as WebPPL
\cite{dippl}, Anglican \cite{wood2014a}, and Birch \cite{murray2018delayed}.
Both WebPPL and Anglican are higher-order, functional \glspl{ppl}, while Birch
is an imperative, object-oriented \gls{ppl}. Anglican includes many \gls{smc}
algorithms, including different variations of \emph{particle \gls{mcmc}
}~\cite{wood2014a}. Anglican also includes various \gls{mcmc} methods.  WebPPL
includes fewer inference algorithms, but both \gls{smc} and \gls{mcmc} methods
are available. Birch performs \gls{smc} inference in combination with using
closed-form optimizations at runtime, automatically yielding a more optimized
version of \gls{smc} taking advantage of \emph{locally-optimal proposals} and
\emph{Rao--Blackwellization}. None of the languages above, however, address the
alignment issue presented in this article. In essence, the programmer needs to
be aware of the internals of the \gls{smc} inference algorithm to write
efficient models---\emph{the model and the inference algorithm have become
coupled}. Optimally, we would like the model and the inference to be as
independent as possible. This is the goal of the work in this paper.



There also exists more theoretical work on \gls{smc} for probabilistic
programming.  One example is a recent denotational validation of \gls{smc} in
probabilistic programming given by Ścibior et
al.~\cite{scibior2017denotational}. This work also includes a denotational
validation of \emph{trace \gls{mcmc}}, another common inference algorithm for
\glspl{ppl}. Trace \gls{mcmc} has also been proven correct by Borgström et. al.
\cite{borgstrom2016a} through an operational semantics for a probabilistic
untyped lambda calculus.


\section{Conclusion}
In this paper, we have introduced an approach for aligning \gls{smc} inference
in \glspl{ppl}. This approach consists of performing a static analysis using
0-CFA, and using this analysis result to automatically align \gls{smc}
inference through a program transformation. We have also evaluated this
approach on a phylogenetic model, showing significant improvements. In
conclusion, we have shown that alignment of \gls{smc} inference in
probabilistic programming can be done automatically, and that it also has a
significant effect on both execution time and accuracy.

\subsection*{Acknowledgements}
This project is financially supported by the Swedish Foundation for Strategic
Research (ASSEMBLE RIT15-0012). We would also like to thank Elias Castegren for
his helpful comments and support.

\appendix

\ifscratchpad
\clearpage
\section{Scratchpad}
\input{scratchpad}
\fi
%
%
%
\clearpage
\bibliographystyle{splncs04}
\bibliography{bibliography}
%
%
%
%
%
\end{document}